\documentclass[aps,prl,twocolumn,showpacs,preprintnumbers,superscriptaddress,nofootinbib]{revtex4-1}
\usepackage{amsmath}
\usepackage{amssymb}
\usepackage{graphicx}
\usepackage{dcolumn}
\usepackage{bm}
\usepackage[usenames,dvipsnames]{color}
\newcommand{\LaMnPOnospace}{\mbox{LaMnPO}}
\newcommand{\LaMnPO}{\mbox{LaMnPO} }

\begin{document}

\title{Antiferromagnetic Exchange, Hund's Coupling and the Origin of the Charge Gap in \LaMnPO}

\author{D.E. McNally}%
\email[]{daniel.mcnally@stonybrook.edu}
\affiliation{Department of Physics and Astronomy, Stony Brook University, Stony Brook, New York 11794-3800, USA}
\author{J. W. Simonson}
\affiliation{Department of Physics and Astronomy, Stony Brook University, Stony Brook, New York 11794-3800, USA}
\author{K. W. Post}
\affiliation{Department of Physics, University of California, San Diego, La Jolla, CA 92093-0319, USA}
\author{Z. P. Yin}
\affiliation{Department of Physics and Astronomy, Rutgers University, Piscataway, NJ 08854, USA}
\author{M. Pezzoli}
\affiliation{Department of Physics and Astronomy, Stony Brook University, Stony Brook, New York 11794-3800, USA}
\affiliation{Department of Physics and Astronomy, Rutgers University, Piscataway, NJ 08854, USA}
\author{G. J. Smith}
\affiliation{Department of Physics and Astronomy,
Stony Brook University, Stony Brook, New York 11794-3800, USA}
\author{V. Leyva}
\affiliation{Department of Physics and Astronomy,
Stony Brook University, Stony Brook, New York 11794-3800, USA}
\author{C. Marques}
\affiliation{Department of Physics and Astronomy,
Stony Brook University, Stony Brook, New York 11794-3800, USA}
\author{L. DeBeer-Schmitt}
\affiliation{Neutron Scattering Sciences Division, Oak Ridge National Laboratory, Oak Ridge TN 37831-6473}
\author{A. I. Kolesnikov}
\affiliation{Neutron Scattering Sciences Division, Oak Ridge National Laboratory, Oak Ridge TN 37831-6473}
\author{Y. Zhao}
\affiliation{NIST Center for Neutron Research, National Institute of Standards and Technology, Gaithersburg, MD 20899, USA}
\author{J.W. Lynn}
\affiliation{NIST Center for Neutron Research, National Institute of Standards and Technology, Gaithersburg, MD 20899, USA}
\author{D. N. Basov}
\affiliation{Department of Physics, University of California, San Diego, La Jolla, CA 92093-0319, USA}
\author{G. Kotliar}
\affiliation{Department of Physics and Astronomy, Rutgers University, Piscataway, NJ 08854, USA}
\author{M. C. Aronson}
\affiliation{Department of Physics and Astronomy, Stony Brook University, Stony Brook, New York 11794-3800, USA}
\affiliation{Condensed Matter Physics and Materials Science Department, Brookhaven National Laboratory, Upton, New York 11973-5000, USA}
\date{\today}

\begin{abstract}

We present inelastic neutron scattering and magnetization measurements of the antiferromagnetic insulator \LaMnPO that are well described by a Heisenberg spin model. These measurements are consistent with the presence of two-dimensional magnetic correlations up to a temperature T$_{max}$ $\approx$ 700 K $\textgreater\textgreater$ T$_{N}$ = 375 K, the N\'eel temperature. Optical transmission measurements show the T = 300 K direct charge gap $\Delta$ = 1 eV has decreased only marginally by 500 K and suggest it decreases by only 10\% at \textit{T$_{max}$}. Density functional theory and dynamical mean field theory calculations reproduce a direct charge gap in paramagnetic \LaMnPO only when a strong Hund's coupling \textit{J$_{H}$} = 0.9 eV is included, as well as onsite Hubbard \textit{U} = 8 eV. These results show the direct charge gap in \LaMnPO is rather insensitive to antiferromagnetic exchange coupling and instead is a result of the local physics governed by \textit{U} and \textit{J$_{H}$}.

\end{abstract}

\pacs{71.3.+h, 74.7.Xa, 75.5.Ee}

\maketitle

The metal-insulator transition in correlated electron systems, where electron states transform from itinerant to localized, has been one of the central themes of condensed matter physics for more than half a century~\cite{imada1998}. In a prototypical Mott transition~\cite{mott1949}, increasing the ratio \textit{U}/\textit{t} of the onsite Hubbard \textit{U} to the kinetic hopping \textit{t} leads to the enhancement of the effective mass of initially itinerant electrons and to spin fluctuations that can  lead to magnetic order. When \textit{U}/\textit{t} surpasses a critical value, the electrons become spatially localized and a metal-insulator transition (MIT) occurs, driven by the formation of a charge gap. Often, the localized electrons are moment-bearing, and magnetic order accompanies the MIT. Thus, electronic localization transitions often involve two different instabilities: magnetic order, involving the spontaneous breaking of rotational symmetry, and a metal-insulator transition that connects an electronic structure with a finite density of states at the Fermi surface to an electronic structure with a charge gap at the Fermi level.

To date, only in select correlated electron materials has it been possible to identify the correlations responsible for the formation of a charge gap. In (V$_{1-x}$Cr$_{x}$)$_{2}$O$_{3}$, the transition from a metal (\textit{x} = 0) to a paramagnetic insulator (\textit{x} = 0.01)  is considered a textbook example of a Mott transition ~\cite{mcwhan1969}. The simultaneous moment collapse and MIT in pressurized MnO results from an increase in the crystal field splitting that eventually overwhelms the strong Hund's coupling \textit{J}$_{H}$ and Hubbard \textit{U} to form a metal~\cite{kunes2008}. In addition to charge gaps arising from local physics (\textit{U} and \textit{J}$_{H}$), magnetic order tends to reinforce electronic localization. Long range magnetic order alone is sufficient to open a gap in a Slater insulator ~\cite{vecchio2013} and even short range magnetic correlations can stabilize the Mott insulating state~\cite{georges1996}.

We describe here a combined experimental and theoretical approach that uses neutron scattering, optical spectroscopy, and electronic structure calculations to determine the origin of the charge gap $\Delta$ = 1 eV in the antiferromagnetic (AF) insulator \LaMnPO\cite{yanagi2009}. While LaMnPO is isostructural to the parent compound of the iron pnictide superconductor LaFeAsO$_{1-x}$F$_{x}$~\cite{hosono2008}, it has a much larger AF ordering temperature \textit{T}$_{N}$ = 375 K and ordered moment $\mu_{AF}$ = 3.2 $\mu_{B}$/Mn, attesting to strong Hund's coupling~\cite{simonson2012}. Hunds coupling is also strong in the metallic parent compounds of the iron superconductors~\cite{haule2009}. However, while Hunds coupling tends to favor the metallic state in the case of Fe, localized Mn systems have a nominal d$^{5}$ configuration in which Hunds coupling is expected to stabilize the charge gap~\cite{georges2013}. Nonetheless, x-ray absorption measurements and electronic structure calculations find significant charge fluctuations in LaMnPO that suggest proximity to a metallic state~\cite{simonson2012}. Metallic LaMnPO was recently realized under pressures of 20 GPa~\cite{guo2013} but so far doping has had only a small effect on the charge gap~\cite{simonson2011}. 

We argue that \LaMnPO is an AF insulator where the inter-atomic exchange interactions \textit{J} play only a limited role in stabilizing the charge gap. Inelastic neutron scattering (INS) measurements at \textit{T} = 5 K $\textless\textless$ \textit{T}$_{N}$ find spin wave excitations extending up to $\approx$ 85 meV. These excitations are well described by a Heisenberg model of interacting local magnetic moments with nearest neighbor exchange \textit{J}$_{1}$ $\approx$ 22 meV and next-nearest neighbor exchange \textit{J}$_{2}$ $\approx$ 7 meV. These exchange interactions suggest a mean field ordering temperature \textit{T}$_{MFT}$ = 760 K $\textgreater\textgreater$ \textit{T}$_{N}$ and high temperature INS measurements are consistent with the presence of antiferromagnetic spin fluctuations up to \textit{T}$_{max}$ $\approx$ 700 K. This is the same temperature where we observe a weak maximum in the measured static and uniform susceptibility $\chi$(\textit{T}), indicating a crossover at \textit{T}$_{max}$ from a state with exchange coupled moments (T $\textless$ T$_{max}$) to a paramagnetic state with individually fluctuating moments (T $\textgreater$ T$_{max}$). Optical transmission measurements show that $\Delta$ decreases slowly as the temperature is increased and suggest that $\Delta$ is suppressed by only $\approx$ 10 $\%$ at \textit{T}$_{max}$, where the correlations become effectively local.

These observations are supported by density functional theory and dynamical mean field theory (DFT+DMFT) calculations in the AF and paramagnetic (PM) states, which show that $\Delta$ only marginally decreases in the absence of AF exchange coupling. Further, DFT+DMFT calculations find a charge gap only when Hunds coupling \textit{J}$_{H}$ is included, along with Hubbard \textit{U}. The strong Hund's coupling that imposes a large fluctuating local moment in the half filled d shell of Mn$^{2+}$ appears to be crucial for the charge gap in \LaMnPOnospace.

Single crystal and polycrystalline \LaMnPO were synthesized as outlined elsewhere~\cite{simonson2011}. Magnetic susceptibility measurements were performed using a Quantum Design Magnetic Property Measurement System and the Vibrating Sample Magnetometer option of a Quantum Design Physical Property Measurement System. Infrared transmission spectra were measured using a Bruker Vertex v/70 FT-IR spectrometer coupled to a custom designed high temperature sample stage.

DFT+DMFT~\cite{DMFT-RMP2006} electronic structure calculations were implemented as in~\cite{Haule-DMFT}, which is based on the full-potential linear augmented plane wave method implemented in Wien2K~\cite{wien2k}. The electronic charge is computed self-consistently on the DFT+DMFT density matrix. The quantum impurity problem is solved by the continuous time quantum Monte Carlo method~\cite{Haule-QMC, werner2006}, using the Slater form of the Coulomb repulsion in its fully rotationally invariant form. We use the experimentally determined crystal structures including the internal positions of the atoms~\cite{nientiedt1997}.

\begin{figure}
\includegraphics[width=9cm]{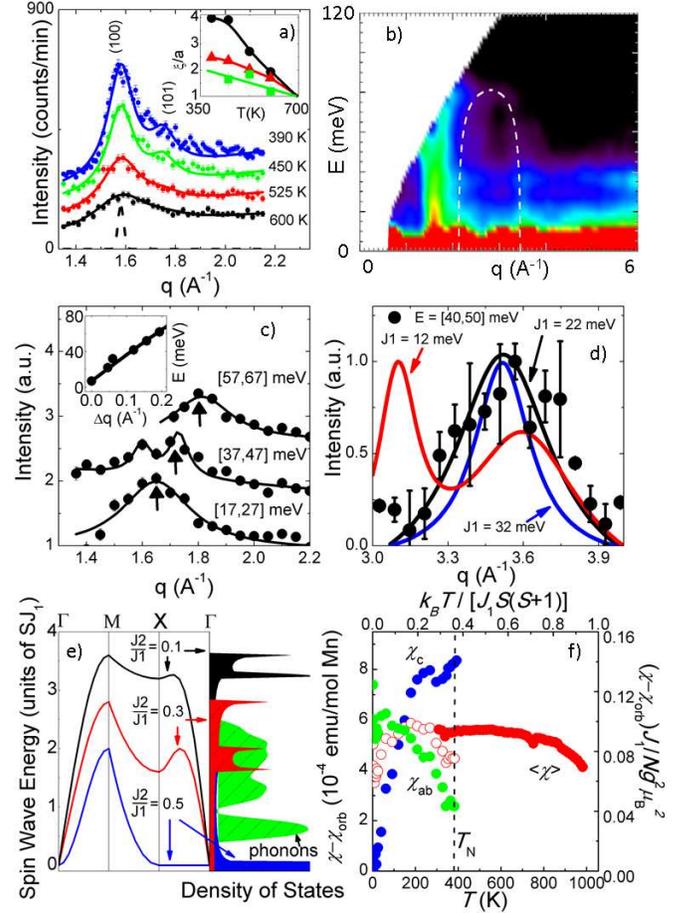}
\caption[]{(Color online)(a) Energy \textit{E} = 5 meV cuts at temperatures \textit{T} indicated. Instrumental resolution is shown as a dashed line. Solid lines are the deconvolutions of the constant \textit{E}-cuts into the sum of two Lorentzians. Inset: \textit{T}-dependence of the spatial correlation length $\xi$=$\Gamma^{-1}$ in units of the lattice constant \emph{a}, for \textit{E} = 5 meV (\textbullet), 10 meV ({\color{red}{$\blacktriangle$}}), 15 meV ({\color{green}$\blacksquare$}). (b) S({\bf{q}},\textit{E}) at 5 K for incident energy of 250 meV. Dashed white line emphasizes the spin wave dispersion $\epsilon$({\bf{q}}) connecting the (110) and (210) zone centers. (c) \textit{E}-cuts near the (100) AF zone center summed over the indicated ranges. Solid lines are fits to the sum of two Lorentzians. Inset: Wave vectors of spin waves $\Delta${\bf{q}}, measured relative to (100), for different \textit{E}. Solid line is theoretical expression for $\epsilon$({\bf{q}}) in $\Gamma$-X direction, with \textit{SJ}$_{1}$ = 34 meV. \textit{S} is total spin, \textit{J}$_{1}$ is the nearest neighbor exchange interaction (d) \textit{E}-cuts near (210) averaged on the interval 40-50 meV. Solid lines represent the theoretical lineshape expected for the powder average of the $\epsilon$({\bf{q}}) for the values of J$_{1}$ indicated. (e) Left: Calculations of $\epsilon$({\bf{q}}) along different directions in reciprocal space for values of \textit{J}$_{2}$/\textit{J}$_{1}$ indicated. \textit{J}$_{2}$ is the next-nearest neighbor exchange interaction. Right: Comparison of the experimental density of states DOS (green shaded area) to the powder average of $\epsilon$({\bf{q}}) for values of \textit{J}$_{2}$/\textit{J}$_{1}$ indicated. The low energy part of the DOS is attributed to phonons. (f) Magnetic susceptibility of a collection of single crystals with field applied in ab plane ($\chi_{ab}$) and c direction ($\chi_{c}$) and the powder average $\langle$$\chi$$\rangle$. Orbital susceptibility $\chi_{orb}$ is subtracted from all data. Dashed line shows \textit{T}$_{N}$ = 375 K.}
\label{neutrons}
\end{figure}

We measured the wavevector q dependence of the scattered neutron intensity S(q) for temperatures \textit{T} $\textgreater$ \textit{T}$_{N}$ = 375 K and for energies \textit{E} $\leq$ 15 meV to look for AF correlations in the PM state. These measurements were performed on a 13 g \LaMnPO powder at the BT-7 triple axis spectrometer at the National Institute of Standards and Technology Center for Neutron Research using a fixed final neutron energy of 14.7 meV~\cite{bt7}.  S(q) at \textit{E} = 5 meV is presented in Fig.~\ref{neutrons}a, along with the instrumental resolution function, showing that the measured peaks are always broader than the resolution. At \textit{T} = 390 K enhanced scattering is found at the q$_{100}$ and q$_{101}$ AF Bragg peaks. With increasing temperature, more spectral weight moves away from these Bragg positions,  suggesting that the longest-lived and longest-range AF correlations are found at \textit{T}$_{N}$, as expected. We fit S(q) with the sum of two Lorentzian functions, as well as a linear background~\cite{reslib}. The centers of the Lorentzian peaks did not vary significantly with temperature or with energy transfers as large as 15 meV. However, the width $\Gamma$ of the peaks was found to increase with increasing temperature. The inverse of the peak width is a measure of the spatial correlation length $\xi$ $\propto$ 1/$\Gamma$, which decreases to a minimum value $\xi$/\textit{a} = 1 near \textit{T}$_{max}$ $\approx$ 700 \emph{K},  as shown in the inset to  Fig.~\ref{neutrons}a. These data demonstrate that the Mn moments are likely effectively decoupled for temperatures above \textit{T}$_{max}$.

We observed high energy spin wave excitations in \LaMnPO and found that these excitations are well described by a Heisenberg \textit{J}$_{1}$-\textit{J}$_{2}$ model. Inelastic neutron scattering measurements with high energy transfers were performed using the SEQUOIA time of flight spectrometer at the Spallation Neutron Source at Oak Ridge National Laboratory~\cite{granroth2010}. A contour plot of the scattered neutron intensity S(q,\textit{E}) at 5 K is presented in Fig. \ref{neutrons}b. There is strong scattering from spin waves at small q and the triple-axis data reveal that there is a spin gap of 7 meV which closes for T $\textgreater$ T$_{N}$. At larger q and \textit{E} $\textless$ 40 meV, S(q,\textit{E}) $\propto$ q$^{2}$, consistent with scattering from phonons. The dashed white line highlights the merging of spin waves originating from the (110) and (210) zone centers where the maximum spin wave energy \textit{E}$_{max}$ $\approx$ 85 meV. Constant energy cuts around the q$_{100}$ Bragg peak are presented in Fig.~\ref{neutrons}c. The peak positions of the Lorentzian fits centered at the larger q are indicated by arrows. The peak positions move to larger q at higher \textit{E}, tracing out the  dispersion of the spin wave excitations. The AF spin wave dispersion for a Heisenberg checkerboard AF is $\epsilon({\bf{q}}) = 4 S J_{1} \sqrt{1-cos^{2}(q_{x} \frac{a}{2}) cos^{2}(q_{y} \frac{a}{2})}$~\cite{zaliznyak2007}, where \textit{S} is the total spin on an atom, \textit{q$_{x}$,q$_{y}$} are the components of {\bf{q}} in the ab plane and \textit{a} is the in plane lattice parameter. The measured dispersion compares favorably to the Heisenberg model for \textit{SJ}$_{1}$=34$\pm$4 meV. Since our sample is polycrystalline, the measured intensity at a given wave vector may include significant contributions from spin waves that originate in different magnetic zones.  Fig.~\ref{neutrons}d compares S(q) at the (210) AF zone center, integrated for energies from 40-50 meV, to the powder averages of the theoretical dispersions for different values of \textit{SJ}$_{1}$. The experimental data are generally consistent with the Heisenberg model for \textit{J}$_{1}$ = 22 meV and \textit{S} = 3/2.

We now examine the spin wave density of states (SWDOS) and find that it is necessary to include \textit{J}$_{2}$ in the Heisenberg model. We determine the ratio \textit{J}$_{2}$/J$_{1}$ by comparing the measured SWDOS with that expected from the Heisenberg model. Spin wave dispersions for key directions in reciprocal space are presented in Fig.~\ref{neutrons}e for values of \textit{J}$_{2}$/\textit{J}$_{1}$ ranging from 0.1-0.5. The corresponding powder averaged SWDOS is compared to the experimentally observed DOS in  Fig.~\ref{neutrons}e. The theoretical SWDOS is most consistent with experiment when 0.2 $\textless$ \textit{J}$_{2}$/\textit{J}$_{1}$ $\textless$ 0.4, yielding a value of 6 meV $\textless$ \textit{SJ}$_{2}$ $\textless$ 14 meV. With these values of exchange interactions,  a mean-field ordering temperature \textit{T}$_{MFT}$ = 4(\textit{J}$_{1}$-\textit{J}$_{2}$)\textit{S}(\textit{S}+1)/(3k$_{B}$) $\geq$ 760 K is expected~\cite{johnston2011}. The reduction of the measured ordering temperature \textit{T}$_{N}$ = 375 K from \textit{T}$_{MFT}$ highlights the quasi-two dimensional nature of \LaMnPOnospace, where the incoherent 2D planes only lock together below \textit{T}$_{N}$=375 K.

\begin{figure}
\includegraphics[width=9cm]{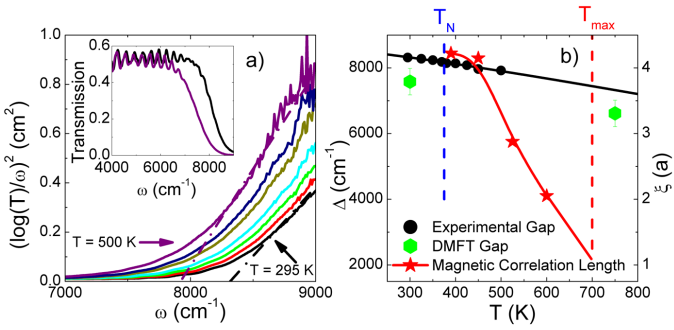}
\caption[]{(Color online) (a) (Log(Transmission)/wavenumber)$^{2}$ for temperatures T = 295 K (\textbf{-}), 325 K ({\color{red}\textbf{-}}), 350 K ({\color{green}\textbf{-}}), 380 K ({\color{cyan}\textbf{-}}), 425 K ({\color{Brown}\textbf{-}}), 450 K ({\color{NavyBlue}\textbf{-}}), 500 K ({\color{Purple}\textbf{-}}). Dashed lines are fits to the 295 K and 500 K data as described in the text. Inset: Raw transmission data for 295 K (black) and 500 K (purple) (b) A comparison of the charge gap $\Delta$ extracted from fits to the optical transmission data, the direct gap in the antiferromagnetic and paramagnetic states determined from DFT+DMFT calculations, and the AF correlation length $\xi$ in units of the in-plane lattice constant \textit{a} extracted from neutron scattering measurements as described in the text.}
\label{optics}
\end{figure}

Indeed, the temperature dependence of the magnetic susceptibility $\chi$(\textit{T}) in \LaMnPO is much as expected~\cite{halperin} for a quasi-2D Heisenberg AF. $\chi$(T) of powdered \LaMnPO crystals in a 1 T field is shown in Fig. \ref{neutrons}f. While no feature is seen at \textit{T}$_{N}$ = 375 K, there is a broad maximum in $\chi$(\textit{T}) centered at $\approx$ 700 K. This is the same temperature where the AF correlations determined from inelastic neutron scattering measurements are projected to vanish. We also measured $\chi$(\textit{T}) between 1.8 K and 400 K on a collection of single crystals with the field oriented along the c-axis ($\chi_{c}$) and with the field in the ab plane ($\chi_{ab}$). The polycrystalline average $\chi$ = 2/3$\chi_{ab}$+1/3$\chi_{c}$.  The normalized \textit{T}=0 susceptibility $\chi_{0}$ = $\chi(\textit{T}=0)$\textit{J}$_{1}$/\textit{Ng}$^{2}\mu_{B}^{2}$~\cite{johnston1997} = 0.063$\pm$0.01 is in good agreement with a modified spin-wave theory~\cite{takahashi1989} with \textit{S} = 3/2 and $J_1$ = 22 meV which yields $\chi_{0}$ = 0.058. The experimental value of the peak susceptibility $\chi_{max}$ = $\chi(T=T_{max})$\textit{J}$_{1}$/Ng$^{2}\mu_{B}^{2}$ = 0.085$\pm$0.05 is in good agreement with the calculated value of 0.091$\pm$0.003, and this modified spin-wave theory provides a very good description of our magnetic susceptibility measurements.

How different is the direct charge gap $\Delta$ in the AF regime than in the PM state with \textit{T} $\geq$ \textit{T}$_{max}$ = 700 \emph{K}? We have measured the optical transmission of a single crystal of \LaMnPO for temperatures as large as 500 K. Raw transmission data for the \LaMnPO sample at 295 K and 500 K are shown in the inset to Fig. \ref{optics}a. With increasing frequency $\omega$, a rapid decrease of the measured transmission is observed, consistent with the onset of absorption due to optical excitations across the energy gap. The main part of Fig. \ref{optics}a presents \mbox{(Log(Transmission)/$\omega$)$^{2}$} for temperatures from 295 K to 500 K. Linear fits to these spectra are used to extract $\Delta$~\cite{rosencher_optoelectronics}, which is plotted as a function of temperature in Fig. \ref{optics}b. The temperature dependence of $\Delta$ was fit using the Varshni equation, which is a simple description of the temperature dependence of $\Delta$ in non-magnetic semiconductors ~\cite{varshni}. This equation accurately describes the data and is indicated by the black line in Fig. \ref{optics}b. The temperature dependence of the magnetic correlation length $\xi$ deduced from our inelastic neutron scattering measurements is also shown in Fig. \ref{optics}b. $\Delta$ is projected to decrease by only $\approx$ 10 \% to 0.9 eV when $\xi$/\textit{a} $\rightarrow$ 1, signalling that the magnetic correlations are confined to the unit cell. 

The picture of \LaMnPO that emerges from our analysis of the measurements presented here is that the Heisenberg exchange interactions \textit{J$_{1}$} and \textit{J$_{2}$} have only a small effect on the magnitude of the direct charge gap $\Delta$. If \textit{J$_{1}$} $\approx$ $\Delta$ then the spin flip energy cost, i.e. the energy cost for an electron to hop between spin-up and spin-down sites, is comparable to $\Delta$ and we would expect the gap to collapse when AF correlations vanish at \textit{T}$_{max}$ = 700 K. However, this is decisively not the case in \LaMnPOnospace, where our  measurements show that \textit{J$_{1}$} $\simeq$ 0.05 $\Delta$ and a sizeable gap remains at \textit{T}$_{max}$, when AF correlations have vanished. Thus, \LaMnPO does not seem to be a Slater-type insulator.

\begin{figure}
\includegraphics[width=8cm]{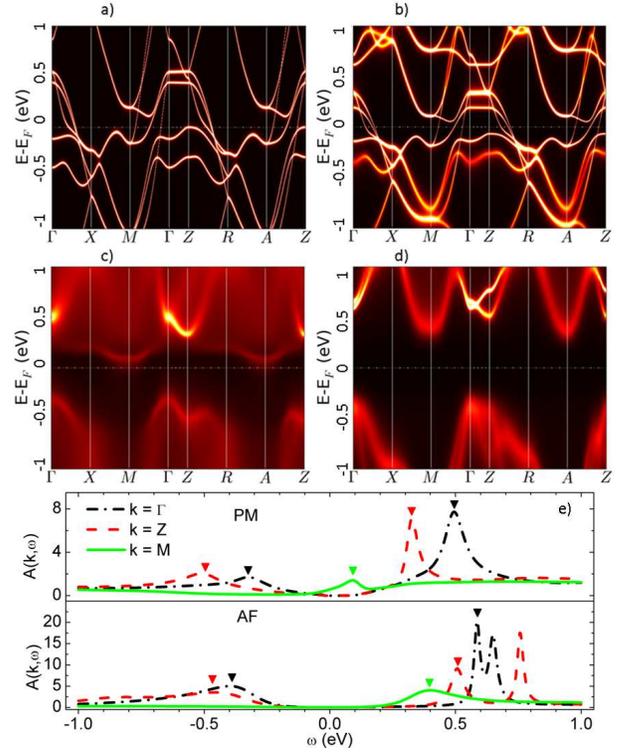}
\caption[]{(Color online)   The theoretical band structure of \LaMnPO   (a) DFT in the non-magnetic state (b) DFT + DMFT in the paramagnetic (PM) state with Hubbard U = 10 eV and Hund's coupling J$_{H}$ = 0 eV (c) DFT + DMFT in the paramagnetic state with U = 8 eV and J$_{H}$ = 0.9 eV (d) DFT + DMFT in the antiferromagnetic (AF) state with U = 8 eV and J$_{H}$ = 0.9 eV (e) DFT + DMFT spectral function A(k,$\omega$) at high symmetry points in the PM state (top) and AF state (bottom). Triangles indicate peak position of A(k,$\omega$).}
\label{dmft}
\end{figure}

We performed electronic structure calculations to clarify the origin of the insulating behavior. A static mean field DFT calculation of the electronic structure of the non-magnetic state, shown in Fig. \ref{dmft}a, predicts that \LaMnPO is metallic, with bands crossing the Fermi level. This result does not agree with our optical transmission measurements which show that \LaMnPO is an insulator with $\Delta$ $\approx$ 1 eV. This suggests that \LaMnPO is not a band insulator and the gap must be a result of electron correlations. We performed DFT+DMFT calculations in the PM state to check whether \LaMnPO can be considered a conventional Mott-Hubbard insulator. Fig \ref{dmft}b  shows the resulting electronic states of \LaMnPO in the presence of a rather large onsite Coulomb interaction \textit{U} = 10 eV~\cite{kotliar2004}. \LaMnPO is still metallic and only moderately correlated with \textit{m}*/\textit{m}$_{band}$ $\approx$ 1.6 for the five Mn 3d orbitals. Thus, we can conclude that the Hubbard U is not solely responsible for the charge gap, and \LaMnPO cannot be considered a conventional Mott-Hubbard insulator.

It has been established that Hund's coupling \textit{J}$_{H}$ is important in multi-band, multi-orbital transition metal systems~\cite{haule2009, yin2011, yin2011_2,georges2011, yin2012}. The first of Hund's empirical rules is that energy is mimimized for a maximum spin \textit{S} on an isolated atom.   For Mn$^{2+}$ ions Hund's rule fills all five 3d orbitals with parallel spins to maximize \textit{S}. This can result in a significant energy cost in hopping between atoms as any doubly occupied orbitals would reduce \textit{S}. Of course, the magnetic moment of \LaMnPO is 3.2 $\mu_{B}$/Mn~\cite{simonson2012}, not 5 $\mu_{B}$/Mn as Hund's rules predict, and a more complete picture of the magnetic correlations in \LaMnPO is required. We performed DFT+DMFT calculations including both Hund's coupling \textit{J}$_{H}$ = 0.9 eV, and Coulomb interaction \textit{U} = 8eV. Fig \ref{dmft}c shows the resulting electronic states in the PM phase of \LaMnPOnospace. \LaMnPO has evolved, by including \textit{J}$_{H}$, from a weakly correlated metal (\textit{U} = $J_{H}$ = 0) to a moderately correlated bad metal(\textit{U} = 10 eV, \textit{J}$_{H}$ = 0), and finally to a bona fide insulator (\textit{U} = 8 eV, \textit{J}$_{H}$ = 0.9 eV). Previously reported DFT+DMFT calculations of AF \LaMnPO are shown in Fig. \ref{dmft}d. The spectral function \textit{A}(\textit{k},$\omega$) at high symmetry points, taken from Fig \ref{dmft}c and Fig \ref{dmft}d, is shown in Fig. \ref{dmft}e. The direct charge gaps, defined from the maxima of \textit{A}(\textit{k},$\omega$), are similar at the $\Gamma$ and Z points with values $\Delta_{AF}$ = 0.94 $\pm$ 0.05 eV and $\Delta_{PM}$ = 0.82 $\pm$ 0.05 eV.   These calculated values are in good agreement with the experimentally determined values of 1 eV in AF state and 0.9 eV in PM state. The indirect gap is defined from the conduction band minimum at M to the valence band maximum at $\Gamma$ as shown in Fig. \ref{dmft}e. While the indirect gap has decreased substantially from 0.74 $\pm$ 0.05 eV in the AF state to 0.4 $\pm$ 0.05 eV in the PM state, it is still much larger than the activation gap $\epsilon$$_{A}$ = 100 meV found in resistivity measurements \cite{simonson2011}, suggesting that the conduction in \LaMnPO is still dominated by in-gap states in the PM phase. These DFT+DMFT calculations support our experimental observation that AF exchange plays a limited role in the formation of a charge gap in LaMnPO and further shows that Hunds coupling is essential for the insulating state.

By combining neutron scattering and optics measurements with electronic structure calculations we have established that \LaMnPO is an example of a correlated insulator where intersite magnetic correlations have little effect on the direct charge gap, which is instead a result of the strong Couloumb interactions \textit{U} and Hund's coupling \textit{J}$_{H}$. Thus, \LaMnPO may be considered as an example of a Mott-Hunds insulator. Superconductivity in the cuprates results from the doping of a Mott insulator, while in the iron pnictides superconductivity results from doping or pressurizing a Hund's metal. It remains to be seen whether doping a Mott-Hund's insulator can lead to a more correlated Hund's metal, where even higher \textit{T}$_{c}$ superconductivity could be expected.

We acknowledge the Office of the Assistant Secretary of Defense for Research and Engineering for providing the NSSEFF funds that supported this research. We acknowledge the support of the National Institute of Standards and Technology, U. S. Department of Commerce in providing neutron research facilities used in this work. Research at the Spallation Neutron Source at Oak Ridge National Laboratory was sponsored by the Scientific User Facilities Division, Office of Basic Energy Sciences, US Department of Energy.

\end{document}